\documentclass[prb,twocolumn,superscriptaddress,showpacs,amsmath,amssymb]{revtex4-2}
\usepackage{graphicx}
\usepackage{amsmath}
\usepackage{amssymb}
\usepackage{float}
\usepackage{color}

\usepackage{cancel,xcolor}
\usepackage{indentfirst}
\usepackage{CJKulem}
\usepackage{soul,xcolor}
\setstcolor{red}
\usepackage{ulem}
\usepackage{cancel}
\usepackage[pdfpagemode=UseNone,pdfstartview=FitH,colorlinks=true,linkcolor=blue,urlcolor=blue,anchorcolor=blue,citecolor=blue]{hyperref}

\begin{document}


\title{Light-tunable quantum metric non-linear Hall response in Berry dipole semimetals}
	\author{Debashree Chowdhury}
	\email{debashreephys@gmail.com}
		\affiliation{Centre for Nanotechnology, Indian Institute of Technology Roorkee, Roorkee, Uttarakhand-247667}
\author{Awadhesh Narayan}
\email{awadhesh@iisc.ac.in}
\affiliation{Solid State and Structural Chemistry Unit, Indian Institute of Science, Bangalore 560012, India}

\begin{abstract}
We investigate the effect of light on quantum metric-mediated intrinsic nonlinear Hall conductivity in Berry dipole semimetals. We discover that light induces a tunable asymmetry in the off-diagonal part of the quantum metric, which is manifested by an asymmetry in the quantum metric dipole. We show that the nonlinear response can be tuned directly by the light amplitude. In particular, we note that the direction of the nonlinear Hall signal changes when the light amplitude is increased beyond a threshold value. Light thus emerges as a promising stimulus to control the quantum geometric response in topological semimetals.
\end{abstract}
\date{\today}

\maketitle\section{Introduction}

Non-linear Hall conductivity (NLHC) arising from non-zero Berry curvature dipoles (BCDs) has recently been an active area of study~\cite{Sodemann,Zhang,Du1,Ortix,Bandyopadhyay}. A non-zero BCD is the first-order momentum derivative of the Berry curvature~\cite{Du,Liao} and requires an inversion symmetry broken system. Besides, the quantum metric is used to measure the distance between the quantum states. The Berry curvature and quantum metric are the imaginary and real parts of the quantum geometric tensor~\cite{Provost,Verma,Gao2,Torma,Liu}. Recently, it has been found that the Berry curvature is not the only driver of NLHC, and quantum metric-based components are a major and yet unexplored source of NLHC~\cite{WangNiu,Ezawa,Das,Yu,Akiba}. In particular, the quantum metric contributes to the NLHC when the quantum metric dipole (QMD) is non-vanishing, which is due to structural symmetries such as parity ($P$) and rotational symmetries.

The search for intrinsic (scattering time independent) second order Hall effects has recently been focusing on the role of quantum metric in MnBi$_{2}$Te$_{4}$/black phosphorus heterostructures~\cite{Gao,Wang}. In these systems the black phosphorus layer needs to be added to remove the $C_{3z}$ rotation of the $PT$-symmetric MnBi$_{2}$Te$_{4}$ of the heterostructure~\cite{Gao}. A large NLHC in antiferromagnetic topological material EuSn$_{2}$As$_{2}$ has also been reported~\cite{Tien}, along with other promising results~\cite{Jin,Dixit,Han}. More recently Yu \textit{et al.} have discussed a higher-order nonlinear intrinsic Hall conductivity in ferromagnetic materials~\cite{Yu1}. Ulrich and co-authors have investigated various aspects of NLHC in several topological systems~\cite{Ulrich}. Inspired by these works, we have studied the nonlinear response in Berry dipole semimetals (BDSs). In BDSs, unlike the Weyl semimetals, the point dipole is formed by two Weyl nodes at a mirror-symmetric interface and thus their mutual annihilation can be avoided~\cite{PRB110,Nelson,Sun,Mo}. Interestingly, in this system the $P$ and $T$ symmetries are broken naturally and the $C_{4z}$ symmetry is preserved.

\begin{figure}[b]
   \centering
   \includegraphics[width=0.9\linewidth]{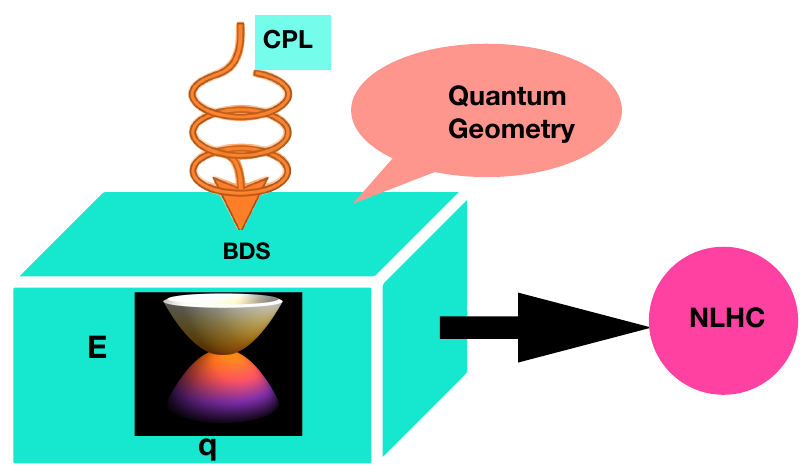}
   \caption{(a) A schematic representation of our studied system. The Berry dipole semimetal (labeled as BDS), shows a quadratically dispersing band structure, in contrast to the linear bands in Weyl or Dirac materials. Here the conduction and valence bands touch each other at a point dipole node. We illuminate the BDS by circularly polarized light and analyze the quantum metric. Using it, we further explore the NLHC of the system in the presence of light.}
   \label{fig:0}
\end{figure}  

The topological properties of a system is possible to tune via light, which provides a window into a prominent field of research~\cite{f1,f2,f3,f4,f5,f6,f7,f8,f9,f10,f11,f12}. This opens a window for generating unique phases~\cite{Oka,Sato,Zhan}. This interaction is particularly compelling in topological semimetals. For example, circularly polarized light can tune the separation between Weyl nodes in Weyl semimetals, which causes renormalized anomalous Hall responses~\cite{Chen1,Chen2}. Berry curvature induced light-tunable linear and nonlinear Hall effects have been well explored, however  circularly polarized light modulated quantum metric-mediated nonlinear responses remain an open area to investigate.

\begin{figure*}[t]
   \centering
   \includegraphics[width=0.95\linewidth]{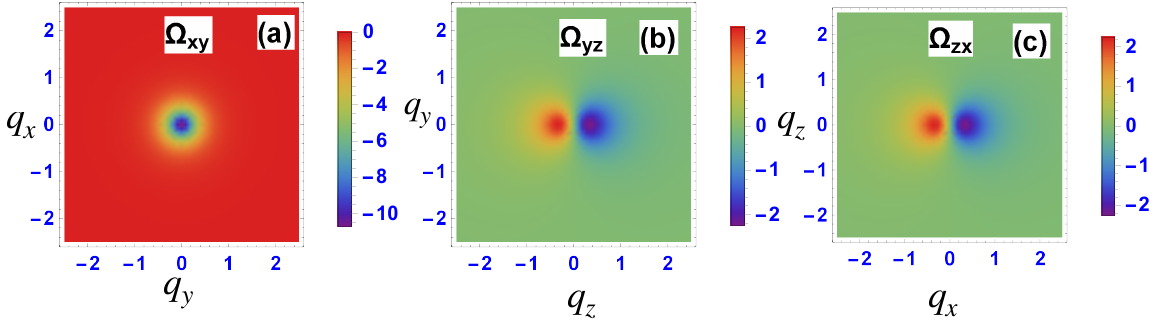}
    \caption{(a) The Berry curvature $\Omega_{xy}$ is plotted with $q_{x}$ and $q_{y}$ keeping $q_{z}=0.5$. $\Omega_{xy}$ does not show the dipole nature. In panel (b), we plot $\Omega_{yz}$ with $q_{z}$ and $q_{y}$ keeping $q_{x}=0.5$. This shows a dipolar nature. This dipole nature is also visible in (c), where we plot $\Omega_{zx}$ as a function of $q_{z}$ and $q_{x}$, keeping $q_{y}= 0.5$. Importantly, in (b) and (c), as we vary $q$, we clearly observe the dipole nature of the Berry curvature.}
   \label{fig:1}
\end{figure*}

In this paper, we characterize the Berry curvature and quantum metric of a BDS. We use these quantities to compute the QMD, which we find to have non-vanishing values. We then study the high-frequency effective Hamiltonian when the system is driven by circular-polarized light and present the quantum metric components. Interestingly, we find that in the absence of light the off-diagonal part of the quantum metric vanishes. However, as we increase the light amplitude, we find an asymmetric off-diagonal part. This asymmetry plays an important role in obtaining a finite NLHC. Beyond a critical light amplitude, the direction of the NLHC reverses. Thus we propose driving as a promising approach to enhance the nonlinear quantum metric Hall response in topological semimetals.

The paper is organized as follows in Sec. II,  we discuss the model for BDSs, and the corresponding Berry curvature. Next we calculate the components of the quantum metric a in Sec. III.  Sec. IV deals with the circularly polarized light and its effect on the quantum metric components. The NLHCs are presented in Sec. V, where we calculate the QMD analytically and discuss the changes light brings in this quantity. Finally we conclude in Sec. VI. \\

\section{Model Hamiltonian and the Berry curvature}

The low-energy Hamiltonian for the first-order BDS is written as~\cite{PRB110}

\begin{align}
H&=(2\chi v v_{z}q_{z}q_{y})\sigma_{x}+2\chi v v_{z}q_{z}q_{x}\sigma_{y}\nonumber\\&+\Big[v^{2}(q_{x}^{2}+q_{y}^{2})-v_{z}^{2}q_{z}^{2}\Big]\sigma_{z},
\end{align}

where $q_{x},q_{y},q_{z}$ are the components of the momentum from the corresponding nodes and $\chi=\pm 1$. Here $v,v_{z}$ are related to the lattice parameters of the model~\cite{PRB110}. The eigenvalues of the Hamiltonian are written as

\begin{align}
	{\cal E}_{\pm}&=\pm\Big[2 q_{x}^{2} q_{y}^{2} v^4+4 q_{x}^{2} q_{z}^{2} v^2 v_{z}^{2} \chi^2-2 q_{x}^{2} q_{z}^{2} v^2 v_{z}^{2}+q_{x}^{4} v^4\nonumber\\&+4 q_{y}^{2} q_{z}^{2} v^2 v_{z}^{2} \chi ^2-2 q_{y}^{2} q_{z}^{2} v^2 v_{z}^{2}+q_{y}^{4} v^{4}+q_{z}^{4} v_{z}^{4}\Big]^{1/2}.
\end{align}

BDSs are the unique topological semimetals where the opposite charge Weyl nodes form a point dipole, which is protected by mirror symmetry. The resulting band structure is thus quadratic. Fig.~\ref{fig:0} shows a schematic of the model considered here, which also presents a schematic band structure of a BDS. The band dispersion is nonlinear, in contrast to the other well-known topological semimetals. The Berry curvature components are calculated as 

\begin{align}
 \Omega^{}_{xy}&=-\frac{2 q_{z}^{2} v^{2}_{} v_{z}^{2}}{\left[\Big(q_{x}^{2}+q_{y}^{2}\Big) v^{2}_{}+q_{z}^{2} v_{z}^{2}\right]^{2}},\nonumber\\
 \Omega^{}_{yz}&=-\frac{2 q_{z}^{}q_{x}^{} v^{2}_{} v_{z}^{2}}{\left[\Big(q_{x}^{2}+q_{y}^{2}\Big) v^{2}_{}+q_{z}^{2} v_{z}^{2}\right]^{2}},\nonumber\\
 \Omega^{}_{zx}&=-\frac{2 q_{z}^{}q_{y}^{} v^{2}_{} v_{z}^{2}}{\left[\Big(q_{x}^{2}+q_{y}^{2}\Big) v^{2}_{}+q_{z}^{2} v_{z}^{2}\right]^{2}}.
\end{align}

From the three components of the Berry curvature, we find that some of them behave in a dipole-like manner as a function of momentum. In Fig.~\ref{fig:1}(a) we plot the Berry curvature component $\Omega_{xy}$ for a fixed value of $q_z$ and change the values of $q_x$ and $q_y$. For $\Omega_{yz}$ [Fig.~\ref{fig:1}(b)] and $\Omega_{zx}$ [Fig.~\ref{fig:1}(c)], we fix $q_x$ and $q_y$ respectively. In Ref.~\cite{PRB110}, the authors have shown that the shape of the Berry curvature in these systems mimics a dipole, in contrast to a monopole as in a Weyl semimetal. This is indeed the case in Fig.~\ref{fig:1}.

Besides the Berry curvature, the quantum metric is a less known but a fundamental quantity with a deep connection to nonlinear transport. The components of the quantum metric are described in the next section.

\begin{figure}[t]
   \centering
   \includegraphics[width=0.99\linewidth]{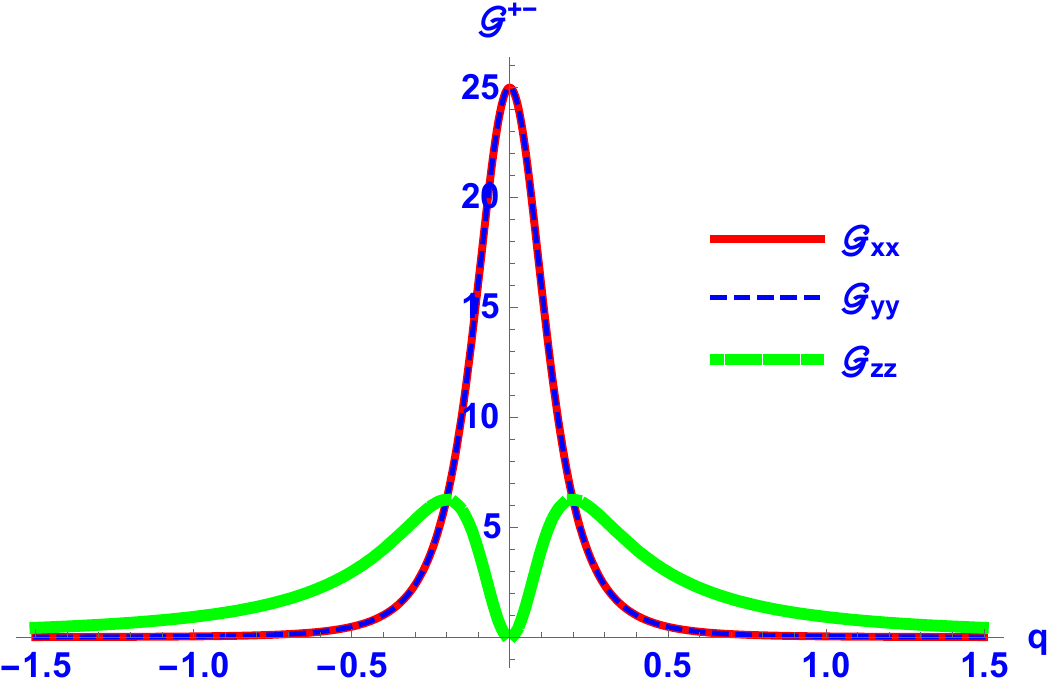}
    \caption{Components of the quantum metric, ${\cal G}_{xx}$ (solid blue), ${\cal G}_{yy}$ (red dashed), and ${\cal G}_{zz}$ (green dot dashed) are plotted with $q=\sqrt{q_{x}^{2}+q_{y}^{2}}$, setting $q_{z}=0.2$. The two diagonal components ${\cal G}_{xx}={\cal G}_{yy}$ show a peak at $q=0$, while ${\cal G}_{zz}$ shows a dip at $q=0$.}
   \label{fig:2}
\end{figure}

\section{Quantum Metric} 

In a Dirac model, the quantum metric components are defined in terms of different components of the Hamiltonian. For a Hamiltonian $H(k)=\mathbf{d}\cdot\mathbf{\sigma}=d_{0}\sigma_{0}+d_{1}\sigma_{x}+d_{2}\sigma_{y}+d_{3}\sigma_{z},$ where $\sigma_{x,y,z}$ and $\sigma_0$ are the Pauli matrices and the identity matrix respectively, the quantum metric components are~\cite{Ezawa2024},

\begin{align}\label{G}
&{\cal G}^{\pm}_{ab}=\frac{1}{4}{\rm Re}\Big[\partial_{k_{a}}\hat{d}\cdot \partial_{k_{b}}\hat{d}\Big],
\end{align}

where $\hat{d}=\left[\frac{d_{x}}{d},\frac{d_{y}}{d},\frac{d_{z}}{d}\right]$ and $d=\sqrt{d_{x}^{2}+d_{y}^{2}+d_{z}^{2}}$. Here $a,b\in\{x,y,z\}$. The superscript $\pm$ indicates the upper or lower band in the two-band model. Using Eq.~(\ref{G}), the diagonal components of the quantum metric are obtained as

\begin{align}
{\cal G}^{}_{xx}&=
\frac{q_{z}^{2} v^2 v_{z}^{2}}{\left( q_{x}^{2} v^{2}_{}+ q_{y}^{2} v^2+q_{z}^{2} v_{z}^{2}\right)^2}=
{\cal G}^{}_{yy},\nonumber\\{\cal G}^{}_{zz}&=\frac{(q_{x}^{2}+q_{y}^{2}) v^{2}_{} v_{z}^{2}}{\left( q_{x}^{2} v^{2}_{}+ q_{y}^{2} v^{2}_{}+q_{z}^{2} v_{z}^{2}\right)^2}.
\end{align}

Interestingly, the component ${\cal G}_{zz}$ shows a double peak pattern (see Fig.~\ref{fig:2}). The off-diagonal component ${\cal G}^{}_{xy}$ is zero in this case. We note that the $xx$ and $yy$ components of the quantum metric are the same, while the $zz$ component is not only distinct, but also contains two parts proportional to $q_{x}^{2}$ and $q_{y}^{2}$ respectively. This results in the underlying structural difference of the diagonal components (Fig.~\ref{fig:2}).

\begin{figure*}[t]
   \centering
    \includegraphics[width=0.95\linewidth]{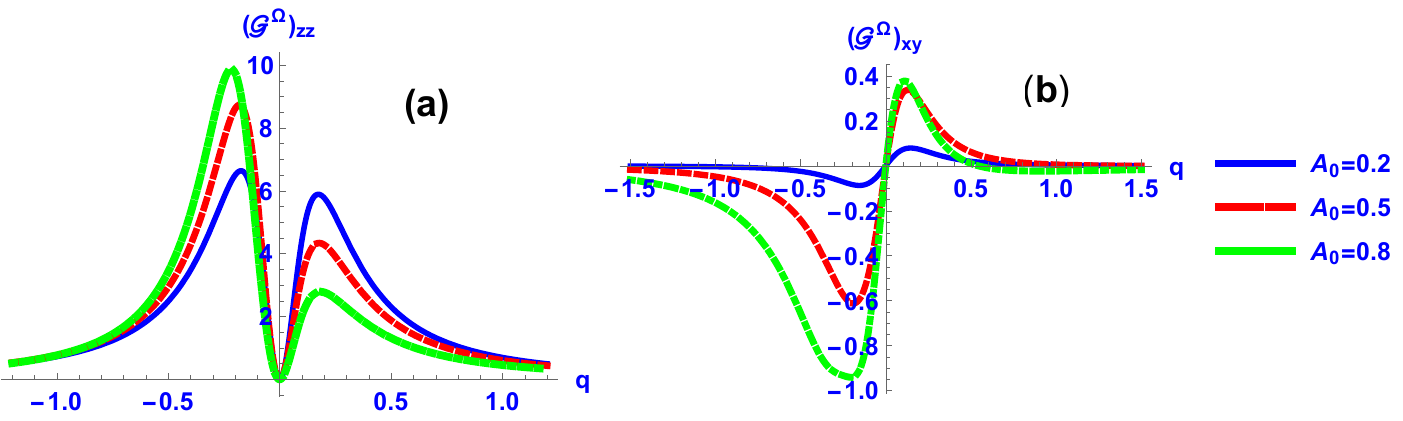}
    \caption{(a) The diagonal component of the quantum metric, ${\cal G}_{zz}$, plotted with $q=\sqrt{q_{x}^{2}+q_{y}^{2}}$ for $A_{0}=0.2$,  $A_{0}=0.5$, and $A_{0}=0.8$, taking $q_{z}=0.3$. (b) The off-diagonal component of the quantum metric, ${\cal G}_{xy}^{\Omega}$, plotted as a function of $q=\sqrt{q_{x}^{2}+q_{y}^{2}}$ for $q_{z}=0.3$. We note that increasing light amplitude increases the peak asymmetries in (b). This asymmetry is crucial for the emergence of the NLHC.}
   \label{fig:3}
\end{figure*}  

We find that although the $xy$ component of the quantum metric vanishes, the Berry curvature is not zero. In the next section, we consider how this situation plays a crucial role in presence of circularly polarized light.

\section{Quantum metric in presence of circular polarized light}

Symmetries are fundamental to the BDS, in which both inversion ($P$) and time-reversal ($T$) symmetries are broken individually. The BDS Hamiltonian initially has $C_{4z}$ rotational symmetry, and when $C_{2v}$ is lifted, we have new nonlinear Hall phenomena. To break this rotational symmetry, we introduce circularly polarized light. In particular, we explore the high-frequency optical driving and explore the possibilities of a non-zero NLHC component to emerge.

We begin by considering light polarized in the $yz$ plane. The vector potential reads

\begin{align}
    A_{y}(t)=A_{0}\sin(\Omega t+\phi_{1}),~ A_{z}(t)=A_{0}\cos\Omega t.
\end{align}

Here $A_{0}$, $\Omega$, and $\phi_{1}$ are the amplitude, frequency and phase difference between the $y$ and $z$ components, respectively. Using the Floquet-Magnus expansion under the high-frequency approximation~\cite{Floquet}, we find the effective Hamiltonian as,

\begin{align}
    H_{\mathrm{eff}}&=\left[2\chi v v_{z}q_{z}q_{y}+\Delta_{1}\right]\sigma_{x}+\left[2\chi v v_{z}q_{z}q_{x}+\Delta_{2}\right]\sigma_{y}\nonumber\\&+\left[v^{2}(q_{x}^{2}+q_{y}^{2})-v_{z}^{2}q_{z}^{2}+\Delta_{3} \right]\sigma_{z}.
\end{align}

Here, the different $\Delta$ terms take the form

\begin{align}\label{delta}
    \Delta_{1}&=-\frac{2 A_{0}^2 q_{x} q_{y} v^3 v_{z} \chi  \cos (\phi_{1})}{\Omega },\nonumber\\
     \Delta_{2}&=\frac{2 A_{0}^2 v v_{z} \chi  \cos (\phi_{1}) \left(q_{y}^2 v^2+q_{z}^2 v_{z}^2\right)}{\Omega },\nonumber\\
      \Delta_{3}&=\frac{2 A_{0}^2 q_{x} q_{z} v^2 v_{z}^2 \chi ^2 \cos (\phi_{1})}{\Omega }.
\end{align}

This illumination breaks the rotational symmetry of the system in the $q_{z}=0$ plane, a feature directly verified by reversing the sign of $q_{z}$. Next, we examine the evolution of the quantum metric components and the QMD under the influence of circularly polarized light. The diagonal and off-diagonal components of the quantum metric in the presence of light are presented in Appendix~\ref{A}.

Figure~\ref{fig:3}(a) displays the quantum metric component ${\cal G}^{\Omega}_{zz}$ for three distinct light amplitudes. As $A_{0}$ increases, a clear asymmetry emerges in the profile. We note here that since our analysis is based in the high-frequency regime, setting $\Omega = 0$ directly in the analytical expressions does not yield the equilibrium (no-light) results. Instead, the zero-light limit is correctly recovered by setting the light amplitude $A_{0} = 0$ in Eq.~\eqref{dia}.

\begin{figure*}[t]
   \centering
    \includegraphics[width=0.8\linewidth]{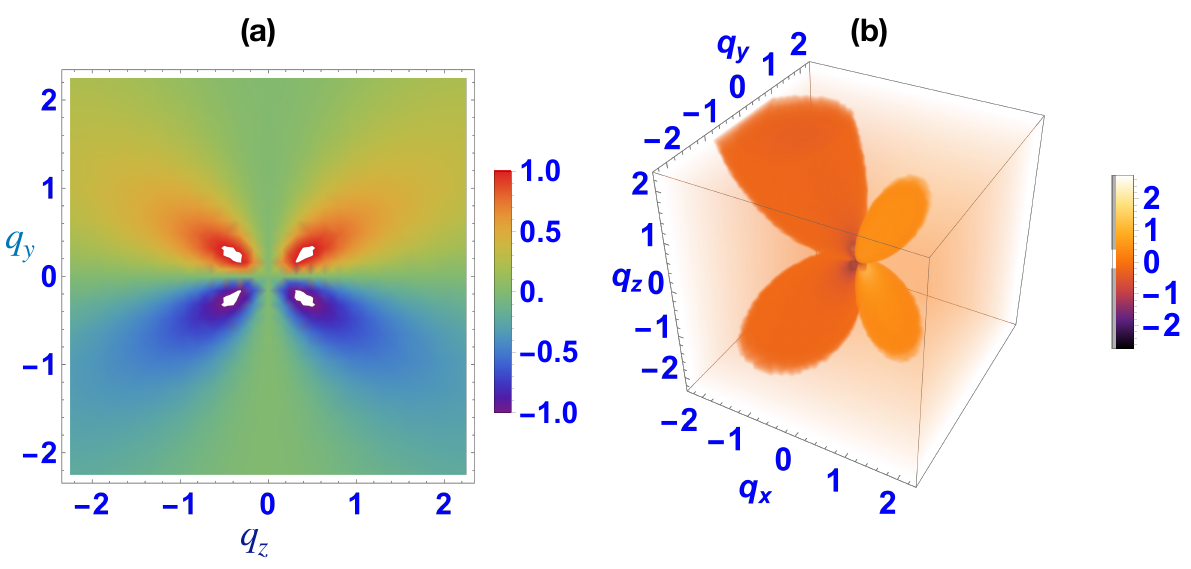}
    \caption{(a) Plot of the integrand $v_{y}{\cal G}_{xx}-v_{x}{\cal G}_{yx}$ with $q_{y}$ and $q_{z}$ taking $q_{x}=0.23$. (b) The same integrand, $(v_{y}{\cal G}^{}_{xx}-v_{x}{\cal G}^{}_{yx})$, is plotted in three-dimensions as a function of $q_{x}$, $q_{y}$, and $q_{z}$. In panel (a), it is evident that although the patterns look symmetric, they are not identical, which is discernible from the central white structures. In panel (b), the asymmetric pattern is clearly visible in the distribution around the $q_{z}=0$ plane.}
   \label{fig:4}
\end{figure*}

Furthermore, in contrast to the pristine BDS case, application of light leads to a non-zero off-diagonal component of the quantum metric along $xy$.

Fig.~\ref{fig:3}(b) illustrates the behavior of ${\cal G}^{\Omega}_{xy}$ as a function of the momentum $q = \sqrt{q_{x}^{2} + q_{y}^{2}}$ for three different representative light amplitudes. Notably, higher light amplitude enhances the asymmetry of the peaks. While the positive values converge to similar values for larger light amplitudes, the negative peaks become more prominent. This feature plays a fundamental role in the emergence of a non-vanishing NLHC, as we will see next. In the following section, we evaluate the QMD for our model, demonstrating how its non-vanishing nature significantly enriches the nonlinear Hall physics of the system.

\begin{figure*}[t]
   \centering
    \includegraphics[width=0.95\linewidth]{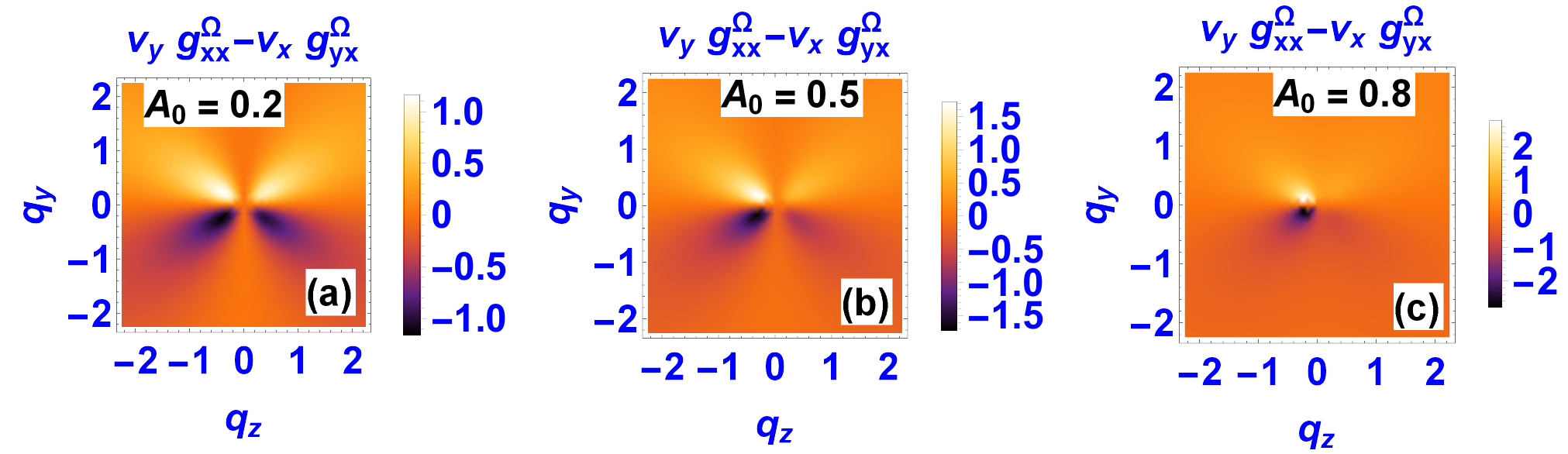}
    \caption{The integrand $(v_{y}{\cal G}^{\Omega}_{xx}-v_{x}{\cal G}^{\Omega}_{yx})$ plotted as a function of $q_{y}$ and $q_{z}$, with $q_{x}=0.23$, for (a) $A_{0}=0.2$, (b) $A_{0}=0.5$, and (c) $A_{0}=0.8$. Note that the plot is asymmetric around the $q_{z}=0$ axis as the light amplitude is increased. We note a shift of the dipole towards negative values of $q_{z}$. This asymmetry leads to a second-order non-linear conductivity.}
   \label{fig:5}
\end{figure*}

\begin{figure*}
   \centering
   \includegraphics[width=0.99\linewidth]{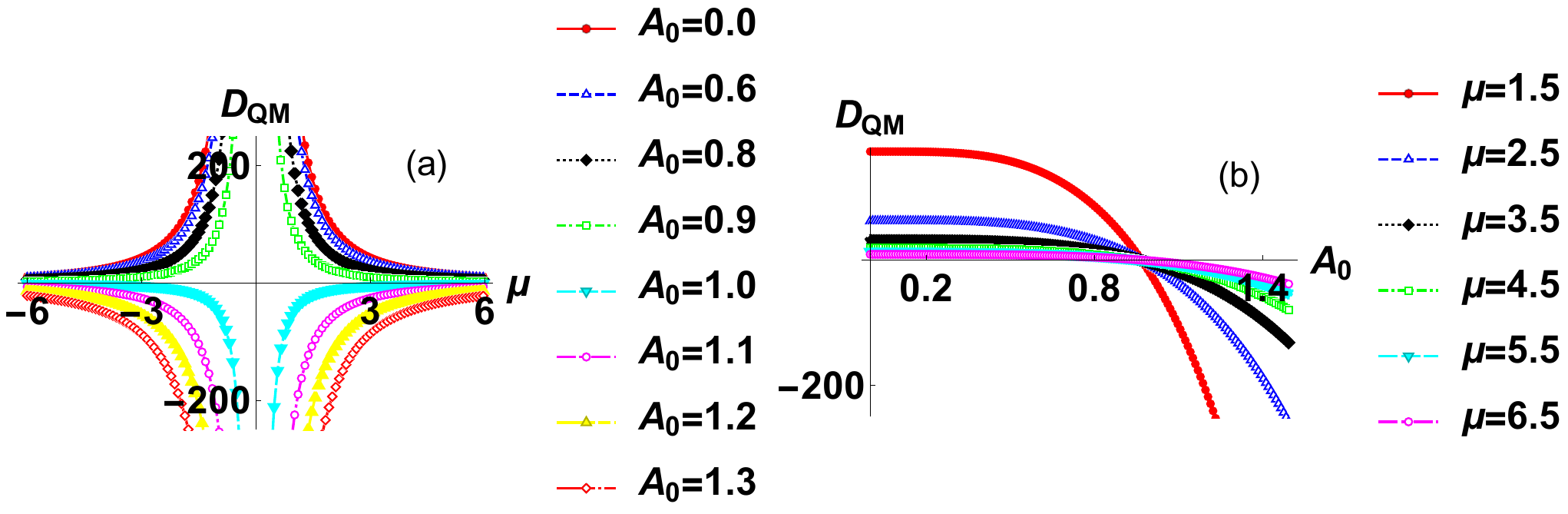}
    \caption{(a) $D_{QM}$ as a function of chemical potential $\mu$ for the BDS with  $A_{0}=0$ (red) $A_{0}=0.6$ (blue), $A_{0}=0.8$ (black), $A_{0}=0.9$ (green), $A_{0}=1.0$ (cyan), $A_{0}=1.1$ (magenta), $A_{0}=1.2$ (yellow) and $A_{0}=1.1$ (orange). With increasing light amplitude, we have a finite value of $D_{QM}$, which decreases for smaller light amplitude. As we increase the light amplitude further, $D_{QM}$ changes sign and increases for higher values of light amplitude beyond a cut-off amplitude ($A_{0}=0.9$). (c) Plot of $D_{QM}$ with increasing $A_{0}$, for different values of $\mu$. We note that below $A_{0}=0.9$, $D_{QM}$ is positive and it has larger value for small $\mu$. However, as we increase $A_{0}$ further, the sign changes and one obtains negative values for $D_{QM}$ with larger values for larger $\mu$.}
   \label{fig:6}
\end{figure*}

\section{NLHC from quantum metric dipole}

In this section, we calculate the QMD for our model, which can be obtained as~\cite{Gao}, 

\begin{align}\label{14}
D_{QM}=\int (v_{y}{\cal G}_{xx}-v_{x}{\cal G}_{yx})\delta(\epsilon-\mu) d\mathbf{q},
\end{align}

where $\mu$ and $v_{x,y}$ are the chemical potential and derivative of the conduction band energy with $q_{x,y}$, respectively. We choose $\epsilon$ as the conduction band energy. Before computing the integral, let us analyze the kernel of the integral $[(v_{y}{\cal G}_{xx}-v_{x}{\cal G}_{xy})]$. The analytical expression for the integrand (without light) is obtained as

\begin{align}
&(v_{y}{\cal G}_{xx}-v_{x}{\cal G}_{xy})=\frac{2 q_{y}^{} q_{z}^{2} v_{}^{4} v_{z}^{2}}{\left(q_{x}^{2} v_{}^{2}+q_{y}^{2} v_{}^{2}+q_{z}^{2} v_{z}^{2}\right)^2}.
\end{align}

Fig.~\ref{fig:4}(a) shows the variation of integrand with $q_{y}$ and $q_{z}$ for fixed $q_{x}$. Although the distribution is mostly symmetric, small differences are seen in the central white areas. Note that if we vary all three momentum components a distinctly asymmetric pattern is achieved [in Fig.~\ref{fig:4}(b)]. The QMD is related to the NLHC through the following relation~\cite{Gao}

\begin{align}
\sigma_{yxx}^{n}=-2 e^{3}\Big[\sum_{n}\frac{D_{QM}^{n}}{\epsilon_{n}-\epsilon_{n'}}\Big]+ \mathrm{AIC},
\end{align}

where AIC are the additional interband contributions~\cite{Gao}, which are not the focus of the current work. To calculate $D_{QM},$ we first discuss the Dirac delta term $\delta(\epsilon_{+}-\mu)$. Considering the conduction band energy and using~\cite{Ezawa2024,Ulrich},

\begin{align}\label{11}
\delta(g(x))=\sum_{i}\frac{\delta(x-x_{i}^{})}{|g'^{}_{}(x_{i})|},
\end{align}

we calculate the real root of the function within the delta function (for $\mu>0$) as

\begin{align}
q_{y,0(1)}^{}&=\sqrt{\pm\mu -v^{2}_{}q_{x}^{2}-v^{2}_{z}q_{z}^{2}},\nonumber\\ q_{y,2(3)}^{}&=-\sqrt{\pm\mu -v^{2}_{}q_{x}^{2}-v^{2}_{z}q_{z}^{2}}.
\end{align}

Importantly in Eq. (\ref{11}), $i$ denotes the roots and here $0,1,2,3$ indicate the four roots of the expression within the $\delta$ function. Once we have obtained these four roots, it is straightforward to compute the other two integrals. We next perform a similar analysis for $D_{QM}$ in the presence of light (see details in Appendix~\ref{B}). The exact expression of the kernel is presented in Eq. (\ref{20}).
In Fig.~\ref{fig:5}, we show the variation of $(v_{y}{\cal G}^{\Omega}_{xx}-v_{x}{\cal G}^{\Omega}_{yx})$ with $q_{y}$ and $q_{z}$ (fixing $q_x$) for increasing values of the light amplitude $A_0$. We find a shift of the dipole towards negative values of $q_{z}$. Overall, with the circular polarized light, the quantum metric exhibits an asymmetry about the $q_{z}=0$ plane. This asymmetry leads to a finite and tunable NLHC.

Following the procedure in Eq. (\ref{14}), we calculate the $\delta$-function and its roots for the system in the presence of light. The analytical results are cumbersome and are omitted for brevity. We provide numerical plots after the threefold integrals. The QMD as a function of the chemical potential $\mu$ under circularly polarized light is presented in Fig.~\ref{fig:6}. As shown in Fig.~\ref{fig:6}(a), the QMD remains non-zero even in the equilibrium case ($A_{0}=0$). However, the effect of light is significant. On the other hand below a critical amplitude (for example, $A_{0} = 0.9$) the dipole can still be seen to have the same qualitative features and orientation as the no-light case. However, at strong values, higher than $A_{0} = 0.9,$ we observe a reversal of the sign of $D_{QM}$, and its magnitude increases to the equilibrium value. This directional flipping is a direct consequence of the light-induced asymmetry in ${\cal G}_{xy}$ as $A_{0}$ increases. Fig.~\ref{fig:6} (b) shows a plot of $D_{QM}$ with $A_{0}$ for different values of the chemical potential $\mu$. While below $\mu=1.5$ we encounter a divergence, above that value, we find a switch in the direction of the QMD, with increasing light amplitude. Thus, light allows direct control over the amplitude as well as the sign of the QMD.

\section{Conclusions}

In summary, we have investigated the effects of light on the quantum metric-driven nonlinear Hall response in an emerging class of topological semimetals, namely, BDSs. We characterize the quantum metric and its modulation in presence of applied circularly polarized light. The interesting finding in this regard is the emergence of an off-diagonal component of the quantum metric that exhibits a finite asymmetry when light is switched on, which in turn produces finite NLHC. Strikingly, the NLHC reverses direction beyond a threshold light amplitude. Overall, our results reveal light to be an interesting approach to tune quantum metric-induced nonlinear Hall response in topological semimetals.

Finally, we address the experimental feasibility of our proposal. Our theoretical predictions could be realized in topological semimetals having higher-order multipole features. In a multi-terminal Hall-bar setup, the NLHC can be measured with an ultrafast pulse to introduce the light-induced asymmetric component.  The switching of the nonlinear Hall current direction with the light amplitude is the key finding of our paper. This light-induced sign flip provides a robust mechanism to distinguish intrinsic QMD effects from extrinsic contributions.\\

\begin{acknowledgments}
D.C. acknowledges financial support from DST (project number DST/WISE-PDF/PM40/2023). A.N. acknowledges support from DST Core Research Grant (CRG/2023/000114). 
\end{acknowledgments}

\appendix
\section{Quantum metric in presence of circularly polarized light}\label{A}

The quantum metric components in presence of light are presented in this section. We obtain

\begin{widetext}
\begin{align}\label{dia}
&{\cal G}^{\Omega}_{xx}=
 \frac{{\cal L}_{xx}}{\left(q_{x}^2 v^2+q_{y}^2 v^2+q_{z}^2 v_{z}^2\right)^2 \left(4 A_{0}^2 q_{x} q_{z} v^2 v_{z}^2 \Omega +4 A_{0}^4 \left(q_{y}^2 v^4 v_{z}^2+q_{z}^2 v^2 v_{z}^4\right)+\Omega ^2 \left(q_{x}^2 v^2+q_{y}^2 v^2+q_{z}^2 v_{z}^2\right)\right)^2}, \nonumber\\
&{\cal G}^{\Omega}_{yy}=\frac{{\cal L}_{yy}}{\left(v^2 \left(q_{x}^2+q_{y}^2\right)+q_{z}^2 v_{z}^2\right)^2 \left(4 A_{0}^2 q_{x} q_{z} v^2 v_{z}^2 \Omega +4 A_{0}^4 \left(q_{y}^2 v^4 v_{z}^2+q_{z}^2 v^2 v_{z}^4\right)+\Omega ^2 \left(v^2 \left(q_{x}^2+q_{y}^2\right)+q_{z}^2 v_{z}^2\right)\right)^2},\nonumber\\
&{\cal G}^{\Omega}_{zz}=\frac{{\cal L}_{zz}}{\left(v^2 \left(q_{x}^2+q_{y}^2\right)+q_{z}^2 v_{z}^2\right)^2 \left(4 A_{0}^2 q_{x} q_{z} v^2 v_{z}^2 \Omega +4 A_{0}^4 \left(q_{y}^2 v^4 v_{z}^2+q_{z}^2 v^2 v_{z}^4\right)+\Omega ^2 \left(v^2 \left(q_{x}^2+q_{y}^2\right)+q_{z}^2 v_{z}^2\right)\right)^2},  
\end{align}

where

\begin{align}
    {\cal L}_{xx}&=v^2 v_{z}^2 \Big(A_{0}^4 v^2 \Omega ^2 \left(2 q_{x}^2 \left(5 q_{y}^2 q_{z}^2 v^4 v_{z}^2+q_{y}^4 v^6+4 q_{z}^4 v^2 v_{z}^4\right)+q_{x}^4 q_{y}^2 v^6+\left(q_{y}^2 v^2+q_{z}^2 v_{z}^2\right)^2 \left(q_{y}^2 v^2+4 q_{z}^2 v_{z}^2\right)\right)\nonumber\\&+4 A_{0}^2 q_{x} q_{z}^3 v^2 v_{z}^2 \Omega ^3 \left(q_{x}^2 v^2+q_{y}^2 v^2+q_{z}^2 v_{z}^2\right)+8 A_{0}^6 q_{x} q_{z} v^4 \Omega  \left(q_{y}^2 v^2 v_{z}+q_{z}^2 v_{z}^3\right)^2\nonumber\\&+4 A_{0}^8 v^4 v_{z}^2 \left(q_{y}^2 v^2+q_{z}^2 v_{z}^2\right)^3+q_{z}^2 \Omega ^4 \left(q_{x}^2 v^2+q_{y}^2 v^2+q_{z}^2 v_{z}^2\right)^2\Big),\nonumber\\
    {\cal L}_{yy}&=v^2 v_{z}^2 \left(A_{0}^2 q_{x} v^2-q_{z} \Omega \right)^2 \Big(4 A_{0}^2 q_{x} q_{z} v^2 v_{z}^2 \Omega  \left(v^2 \left(q_{x}^2+q_{y}^2\right)+q_{z}^2 v_{z}^2\right)\nonumber\\&+4 A_{0}^4 v^2 v_{z}^2 \left(q_{z}^2 v^2 v_{z}^2 \left(q_{x}^2+2 q_{y}^2\right)+q_{y}^4 v^4+q_{z}^4 v_{z}^4\right)+\Omega ^2 \left(v^2 \left(q_{x}^2+q_{y}^2\right)+q_{z}^2 v_{z}^2\right)^2\Big),\nonumber\\
    {\cal L}_{zz}&=v^2 v_{z}^2 \Big(4 A_{0}^4 v^2 v_{z}^2 \Omega ^2 \left(q_{x}^2+q_{y}^2\right) \left(2 q_{z}^2 v^2 v_{z}^2 \left(q_{x}^2+q_{y}^2\right)+q_{y}^2 v^4 \left(q_{x}^2+q_{y}^2\right)+q_{z}^4 v_{z}^4\right)\nonumber\\&+4 A_{0}^2 q_{x} q_{z} v^2 v_{z}^2 \Omega ^3 \left(q_{x}^2+q_{y}^2\right) \left(v^2 \left(q_{x}^2+q_{y}^2\right)+q_{z}^2 v_{z}^2\right)\nonumber\\&+4 A_{0}^8 q_{x}^2 v^4 v_{z}^4 \left(q_{x}^2 q_{y}^2 v^4+\left(q_{y}^2 v^2+q_{z}^2 v_{z}^2\right)^2\right)+8 A_{0}^6 q_{x}^3 q_{z}^3 v^4 v_{z}^6 \Omega +\Omega ^4 \left(q_{x}^2+q_{y}^2\right) \left(v^2 \left(q_{x}^2+q_{y}^2\right)+q_{z}^2 v_{z}^2\right)^2\Big).
\end{align}

Here, the superscript $\Omega$ indicates the quantum metric obtained in the presence of light.
The off-diagonal component of quantum metric in presence of light is

\begin{align}
  {\cal G}_{xy}^{\Omega} = -\frac{{\cal L}_{xy}}{\left(v^2 \left(q_{x}^2+q_{y}^2\right)+q_{z}^2 v_{z}^2\right)^2 \left(4 A_{0}^2 q_{x} q_{z} v^2 v_{z}^2 \Omega +4 A_{0}^4 \left(q_{y}^2 v^4 v_{z}^2+q_{z}^2 v^2 v_{z}^4\right)+\Omega ^2 \left(v^2 \left(q_{x}^2+q_{y}^2\right)+q_{z}^2 v_{z}^2\right)\right)^2},
\end{align}

where

\begin{align}
    {\cal L}_{xy}&=A_{0}^2 q_{y} v^4 v_{z}^2 \left(A_{0}^2 q_{x} v^2-q_{z} \Omega \right) \Big(2 A_{0}^2 q_{x} q_{z} v^2 v_{z}^2 \Omega  \left(v^2 \left(q_{x}^2+3 q_{y}^2\right)+3 q_{z}^2 v_{z}^2\right)\nonumber\\&+4 A_{0}^4 \left(q_{y}^2 v^3 v_{z}+q_{z}^2 v v_{z}^3\right)^2+\Omega ^2 \left(v^2 \left(q_{x}^2+q_{y}^2\right)+q_{z}^2 v_{z}^2\right)^2\Big).
\end{align}

\section{Quantum metric dipole in presence of light}\label{B}

The quantum metric dipole in presence of light is obtained as

\begin{align}\label{14}
D_{QM}=\int_{q} d\mathbf{q} (v_{y}{\cal G}^{\Omega}_{xx}-v_{x}{\cal G}^{\Omega}_{yx})\delta(\epsilon-\mu).
\end{align}

As we have all the necessary components of the QMD, we write the integrand as 

\begin{align}\label{20}
&(v_{y}{\cal G}^{\Omega}_{xx}-v_{x}{\cal G}^{\Omega}_{yx})=
\frac{{\cal R}}{\Omega  \left(\left(q_{x}^2 v^2+q_{y}^2 v^2+q_{z}^2 v_{z}^2\right) \left(4 A_{0}^2 q_{x} q_{z} v^2 v_{z}^2 \Omega +4 A_{0}^4 \left(q_{y}^2 v^4 v_{z}^2+q_{z}^2 v^2 v_{z}^4\right)+\Omega ^2 \left(q_{x}^2 v^2+q_{y}^2 v^2+q_{z}^2 v_{z}^2\right)\right)\right)^{3/2}},
\end{align}

where,

\begin{align}
{\cal R}&=2 q_{y} v_{z}^2 \Big(-A_{0}^2 q_{x} q_{z} v^6 \Omega ^3 \left(q_{x}^2 v^2+q_{y}^2 v^2-q_{z}^2 v_{z}^2\right)+A_{0}^4 v^6 \Omega ^2 \Big(2 q_{x}^2 \left(q_{y}^2 v^4+q_{z}^2 v^2 v_{z}^2\right)\nonumber\\&+q_{x}^4 v^4+4 q_{y}^2 q_{z}^2 v^2 v_{z}^2+q_{y}^4 v^4+3 q_{z}^4 v_{z}^4\Big)+A_{0}^6 q_{x} q_{z} v^8 v_{z}^2 \Omega  \left(q_{x}^2 v^2+5 q_{y}^2 v^2+5 q_{z}^2 v_{z}^2\right)\nonumber\\&+4 A_{0}^8 v^8 \left(q_{y}^2 v^2 v_{z}+q_{z}^2 v_{z}^3\right)^2+q_{z}^2 v^4 \Omega ^4 \left(q_{x}^2 v^2+q_{y}^2 v^2+q_{z}^2 v_{z}^2\right)\Big).
\end{align}
\end{widetext}

\end{document}